\documentclass[12pt,english]{article}
\usepackage{amsmath}
\usepackage{amssymb}
\usepackage{mathrsfs}
\usepackage{graphicx}

\makeatletter

\catcode`@=12 \topmargin -0.5in \evensidemargin 0.0in
\usepackage{babel}
\makeatother
\newcommand {\be}{\begin{equation}}
\newcommand {\ee}{\end{equation}}
\newcommand {\ba}{\begin{eqnarray}}
\newcommand {\ea}{\end{eqnarray}}

\begin{document}
\thispagestyle{empty}
\begin{flushright}
IPM/P-2009/016\\
\today
\end{flushright}

\mbox{} \vspace{0.75in}

\begin{center}\textbf{\large An Analysis of  Cosmic Neutrinos: }\\
\textbf{ Flavor Composition at Source  and Neutrino Mixing
Parameters}\\
 \vspace{0.5in} \textbf{$\textrm{Arman Esmaili}^{\dag\S}$\footnote{arman@mail.ipm.ir} and $\textrm{Yasaman Farzan}^{\S}$\footnote{yasaman@theory.ipm.ac.ir}}\\
 \vspace{0.2in} \textsl{${}^\dag$Department of Physics, Sharif University of Technology\\P.O.Box 11365-8639, Tehran, IRAN}\\
 \vspace{0.2in}\textsl{${}^\S$Institute for Research in Fundamental Sciences (IPM)\\P.O.Box 19395-5531, Tehran, IRAN}\\
 \vspace{.55in}\end{center}

\baselineskip 18pt
\begin{abstract}

We examine the feasibility of deriving neutrino mixing parameters
$\delta$ and $\theta_{13}$ from the cosmic neutrino flavor
composition under the assumption that the flavor ratios of the
cosmic neutrinos at the source were
$F_{\nu_e}+F_{\bar{\nu}_e}:F_{\nu_\mu}+F_{\bar{\nu}_\mu}:
F_{\nu_\tau}+F_{\bar{\nu}_\tau}=1:2:0$. We analyze various
uncertainties that enter the derivation of $\delta$ and
$\theta_{13}$ from the ratio of the shower-like to $\mu$-tracks
events which is the only realistic source of information on the
flavor composition at neutrino telescopes such as ICECUBE. We then
examine to what extent the deviation of the initial flavor ratio
from $1:2:0$ can be tested by  measurement of this ratio at
neutrino telescopes taking into account various sources of
uncertainty.

\end{abstract}

\section{Introduction}

According to the current models, the astrophysical objects such as
sources of Gamma Ray Bursts (GRBs) \cite{GRB}, type I b/c
supernovae \cite{SNR} and Active Galactic Nuclei (AGNs) \cite{AGN}
can emit beams of neutrinos luminous enough to be detectable at
the neutrino telescopes that are under construction. The AMANDA
experiment \cite{AMANDA} at the south pole, which came to the end
of its mission in 2006, has set the following bound on the diffuse
flux of neutrinos \be E^2_\nu \frac{dF_\nu}{dE_\nu}\leq 8.2\quad\;
{\text{GeV cm$^{-2}$ sr$^{-1}$ yr$^{-1}$}} \ . \label{BoundONFlux}
\ee A km$^3$-scale neutrino telescope named ICECUBE is under
construction which encompasses AMANDA. If the bound in
(\ref{BoundONFlux}) is saturated, the completed ICECUBE can
collect $\sim4000$ cosmic neutrino signal each year
\cite{icecubesensitivity}. Notice that this bound is on the sum of
the neutrino fluxes from sources at cosmological distances. In
principle, core collapse supernova explosions leading to an
intense neutrino flux detectable at km$^3$-scale neutrino
telescopes can also take place in the close-by galaxies located at
a distance of $\lesssim 10$ Mpc \cite{optimistic,missed}. If such
an explosion is registered during the time that the ICECUBE is in
full swing, ICECUBE can record about a few hundred neutrino events
from a single explosion \cite{missed}. In addition to ICECUBE in
the south pole, three neutrino telescopes NEMO \cite{NEMO},
ANTARES \cite{ANTARES} and NESTOR \cite{NESTOR} are being
constructed in the Mediterranean sea. Moreover, the so-called
KM3NET neutrino telescope \cite{KM3NET} is planned to be
constructed in the Mediterranean sea.

In view of this prospect, extensive studies have been performed on
the possibility of deriving information on the mixing parameters
of neutrinos by studying the flavor ratio of the neutrinos at the
detector \cite{parameters,theta23,Farzan1}. The method is based on
the following argument. Suppose the flavor ratio at the source was
$F_{\nu_e}+F_{\bar{\nu}_e}:F_{\nu_\mu}+F_{\bar{\nu}_\mu}:F_{\nu_\tau}+F_{\bar{\nu}_\tau}
=w_e^0:w_\mu^0:w_\tau^0 \ .$ After propagating the distance
between the source and the detector, the flavor ratio will become
\be \label{RatioATdetector}
F_{\nu_e}+F_{\bar{\nu}_e}:F_{\nu_\mu}+F_{\bar{\nu}_\mu}:F_{\nu_\tau}+F_{\bar{\nu}_\tau}
=\sum_\alpha w_\alpha^0 P_{\alpha e}:\sum_\alpha w_\alpha^0
P_{\alpha \mu}:\sum_\alpha w_\alpha^0 P_{\alpha \tau} \ , \ee
where $P_{\alpha \beta}$ is the probability of $\nu_\alpha \to
\nu_\beta$. Considering the very long distance between the source
and the Earth ({\it i.e.,} $\Delta m_{ij}^2 L/(2E_\nu)\gg 1$), the
oscillatory terms in $P_{\alpha \beta}$ average out \footnote{For
a detailed discussion of the loss of coherence, see \cite{loss}.}:
\be P_{\alpha \beta}\equiv P(\nu_\alpha \to
\nu_\beta)=P(\bar{\nu}_\alpha \to \bar{\nu}_\beta)=\sum_i
|U_{\alpha i}|^2|U_{\beta i}|^2, \label{standard-oscillation}\ee
where $U_{\alpha i}$ are the elements of the neutrino mixing
matrix. Notice that $P_{\alpha \beta}$ is independent of the
neutrino energy $E_\nu$, distance $L$ and the mass square
differences $\Delta m^2_{ij}$.

In a wide range of models that lead to detectable cosmic neutrino
flux, the neutrino production takes place through $\pi^\pm \to
\mu^\pm \stackrel{(-)}{\nu}_\mu$ and subsequently $\mu \to e \nu_e
\nu_\mu$. The flavor ratios at the source are therefore predicted
to be $w_e^0:w_\mu^0:w_\tau^0=1:2:0$. Thus, by measuring the
flavor ratio at Earth, one can in principle derive the absolute
values of the mixing matrix elements which yield information on
the yet-unknown neutrino parameters $\theta_{13}$ and $\delta$ as
well as the deviation of $\theta_{23}$ from $\pi/4$
\cite{parameters,theta23,Farzan1,Farzan2}. It is also suggested to
employ cosmic neutrinos to discriminate between the standard
oscillation scenario and more exotic possibilities
\cite{meloniANDtommy,zhou}.

The flavor identification power of ICECUBE and other neutrino
telescopes is limited. In fact, in the energy range of interest
($100 ~{\rm GeV}<E_\nu<100~{\rm TeV}$), ICECUBE can only
distinguish between shower-like events and the $\mu$-track events.
Each of these two types of events can receive contributions from
different flavors. As discussed in the next section, several input
and assumptions go into derivation of the flavor composition from
the ratio of the shower-like events to the $\mu$-track events.
Lack of knowledge or uncertainty on these input parameters will
lead to uncertainty in derivation of $\theta_{13}$ and $\delta$
from the cosmic neutrinos. To derive information  on the unknown
mixing parameters ($\theta_{13}$ and/or $\delta$) from the ratio,
one should also consider the effect of the uncertainty on the
known neutrino mixing parameters, $\theta_{23}$ and $\theta_{12}$.
A complete treatment of all these effects is missing in the
literature.  In the first part of this paper, we study the
possibility of deriving $\theta_{13}$ and $\delta$ from the ratio
of shower-like events to $\mu$-track events. We take into account
the uncertainty of the aforementioned inputs as well as the
possible uncertainty in the measurement of the ratio itself. In
our analysis, we include contributions to the shower-like and
$\mu$-track events whose effects can be larger than the effect of
$\delta$ and $\theta_{13}$ but their effects have been overlooked
in the literature. We discuss how much the precision of various
input parameters has to be improved in order to make the
measurement of $\theta_{13}$ or $\delta$ by neutrino telescopes a
reality.

Various effects can cause a substantial deviation of the flavor
ratio from $w^0_e:w^0_\mu:w^0_\tau=1:2:0$. In the second part of
the paper, considering the realistic uncertainties in the input,
we discuss to what extent the ratio at the source can be
determined under the assumption that propagation of neutrinos from
the source to detector is simply governed by the standard
oscillation formula in
Eqs.~(\ref{RatioATdetector},\ref{standard-oscillation}).

The paper is organized as follows. In Sect.~\ref{flavor}, the
mechanism for flavor identification at neutrino telescopes is
described. In Sect.~\ref{standard}, the features of the cosmic
neutrino fluxes predicted by the mainstream models are discussed.
In Sect.~\ref{un-impact}, assuming CP conservation, the effects of
possible sources of uncertainties on the derivation of $s_{13}$
are studied. An analysis of derivation of $\delta$ in the presence
of uncertainties is performed in Sect.~\ref{degeneracy}.
Sect.~\ref{flavor-composition} is devoted to discussing various
possible mechanisms through which the initial ratios can deviate
from $1:2:0$. We discuss whether by measuring the flavor ratio on
Earth, it will be possible to differentiate between models. A
summary of the conclusions is provided in Sect.~\ref{conclusions}.

\section{Flavor identification \label{flavor}}

ICECUBE and its Mediterranean counterparts can basically
distinguish only two types of events: 1) shower-like events; 2)
$\mu$-track events. As shown in the seminal work by Beacom {\it et
al.} \cite{Beacom}, one can derive information on the flavor
composition by studying the ratio of the $\mu$-track events to
shower-like events. Let us define
\begin{equation} \label{ratioR} R=\frac{\text {Number
of Muon-track events}}{\text {Number of Shower-like
events}}.\end{equation} There is a threshold energy, $E_{th}$
below which the neutrino cannot be detected by a neutrino
telescope. The value of $E_{th}$ depends on the structure of the
detector and the type of the event (shower-like versus
$\mu$-track). Since the detection of neutrinos coming from above
suffers from the large background from cosmic rays, the neutrino
telescopes mainly focus on upward going neutrinos; {\it i.e.},
neutrinos that pass through the Earth before reaching the detector
(see however, \cite{veto}). The Earth is opaque for neutrinos with
energies higher than $\sim 100~$TeV \cite{Gandhi}. In sum, the
neutrino telescopes mainly study the muon neutrinos in the range
$E_{th}\sim 100$~GeV and $E_\nu^{cut}\sim 100$~TeV. In calculating
the fluxes we set the upper limit of the integration equal to
$E_\nu^{cut}=100$ TeV. By using this value for the upper limit, we
can neglect the attenuation of the neutrino flux crossing the
Earth which depends on the direction of the neutrino
\cite{Kistler}.

Two sources contribute to the $\mu$-track events: (i) Charged
Current (CC) interaction of $\nu_\mu$ or $\bar{\nu}_\mu$ producing
$\mu$ or $\bar{\mu}$; (ii) CC interaction of $\nu_\tau$ and
$\bar{\nu}_\tau$ producing $\tau$ or $\bar{\tau}$ and the
subsequent decay of $\tau$ and $\bar{\tau}$ into $\mu$ and
$\bar{\mu}$. In the literature, the contribution of $\nu_\tau$
(via $\nu_\tau \to \tau \to \mu$) to $\mu$-track events has been
overlooked but to study the effect of $\theta_{13}$, one should
take into account such sub-dominant effects.

The contribution of $\nu_\mu$ and $ \bar{\nu}_\mu$ to the
$\mu$-track events can be estimated as
\begin{equation} \label{muCC1}\rho A
N_A \iint R_\mu(E_\mu,E_{th}^\mu)\frac{dF_{\nu_\mu}}{dE_{\nu_\mu}}
\frac{d\sigma^{CC}}{dE_\mu} dE_{\mu}dE_{\nu_\mu} +[{\rm particle }
\to {\rm antiparticle}]
 ,\end{equation}
where $\rho$, $A$ and $N_A$ are respectively the density of the
medium (ice/water), the effective area of the detector and the
Avogadro number. $R_\mu(E_1,E_2)$ is the muon range which is the
distance traveled in the medium by a muon with energy $E_1$ before
its energy drops below $E_2$. The muon range in ice is given by
\cite{range}
\begin{equation}\label{icerange} R_\mu(E_1,E_2)=(2.6
{\text { Km}})\ln\left[\frac{2+4.2\times 10^{-3}E_1}{2+4.2 \times
10^{-3} E_2}\right] ,\ \end{equation} where both $E_1$ and $E_2$
are in GeV. Finally, $dF_{\nu_\mu}/dE_{\nu_\mu}$ and $\sigma^{CC}$
are respectively the neutrino flux spectral function at the
detector and the cross section of the CC interactions of
$\nu_\mu$.

The contributions from $\nu_\tau$ and $\bar{\nu}_\tau$ to
$\mu$-track events can be estimated as
\begin{equation} B \rho A N_A
\int^{E_{cut}}\int \int_{E_{th}^\mu}
\frac{dF_{\nu_\tau}}{dE_{\nu_\tau}} \frac{d
\sigma^{CC}}{dE_\tau}f\left( E_{\tau} ,E_\mu
\right)R_\mu(E_\mu,E_{th}^\mu) dE_\mu dE_\tau dE_{\nu_\tau}
+(\nu_\tau \to \bar{\nu}_\tau)\ , \label{tauCCmu}
\end{equation} where $B \equiv {\rm Br}(\tau\to \mu \bar{\nu}_\mu
\nu_\tau)=17.8\%$. The function $f(E_\tau, E_\mu)$ in the above
equation is the probability density of the production of a muon
with energy $E_\mu$ in the decay of a $\tau$ lepton with energy
$E_\tau$. That is \be \label{f-defintion}f(E_\tau ,E_\mu) \equiv
\frac{1}{\Gamma}\frac{d \Gamma(\tau(E_\tau) \to \mu(E_\mu)
\bar{\nu}_\mu \nu_\tau)}{ dE_\mu}.\ee The details of the
calculation of $f(E_\tau,E_\mu)$ can be found in
Appendix~\ref{function}.

Three types of events appear as shower: i) the Neutral Current
(NC) interactions of all kinds of neutrinos; ii) the CC
interactions of $\nu_e$ and $\bar{\nu}_e$; iii) the CC
interactions of $\nu_\tau$ ($\bar{\nu_\tau}$) and the subsequent
hadronic decay of $\tau$ ($\bar{\tau}$). Showers from NC
interaction of all three neutrino flavors can be estimated as \ba
\label{shower} \sum_{l=e,\mu,\tau} \rho ALN_A \left[\int^{E_{cut}}
\frac{dF_{\nu_l}}{dE_{\nu_l}}\sigma^{NC}dE_{\nu_l}+
 \int^{E_{cut}}
\frac{dF_{\bar{\nu}_l}}{dE_{\bar{\nu}_l}}\bar{\sigma}^{NC}dE_{\bar{\nu}_l}\right]
, \ea where $L$ is the length of the detector. The rate of the
electromagnetic showers from the CC interactions of $\nu_e$ and
$\bar{\nu}_e$ is \be \label{shower1} \rho ALN_A
\left[\int^{E_{cut}}
\frac{dF_{\nu_e}}{dE_{\nu_e}}\sigma^{CC}dE_{\nu_e}+ \int^{E_{cut}}
\frac{dF_{\bar{\nu}_e}}{dE_{\bar{\nu}_e}}\bar{\sigma}^{CC}dE_{\bar{\nu}_e}\right].
\ee

The rate of showers originated from the CC interaction of
$\nu_\tau$ with the subsequent hadronic decay of the $\tau$ lepton
is \be \label{shower2} (1-B)\rho ALN_A \left[\int^{E_{cut}}
\frac{dF_{\nu_\tau}}{dE_{\nu_\tau}}\sigma^{CC}dE_{\nu_\tau}+
\int^{E_{cut}}
\frac{dF_{\bar{\nu}_\tau}}{dE_{\bar{\nu}_\tau}}\bar{\sigma}^{CC}dE_{\bar{\nu}_\tau}\right].
\ee

Notice that while the shower-like events are given by the length
of the detector, the $\mu$-track events are given by the muon
range [see Eqs.~(\ref{muCC1},\ref{tauCCmu})]. This is because
muons can emit Cherenkov light and trigger the detector even if
they are produced outside the detector but within the range
$R_\mu$. In other words, for the muon detection, the effective
volume is larger than the geometrical volume.

To write down the above formulas, several simplifications have
been made:
\begin{itemize}
\item Obviously, neutrinos entering the detector through different
zenith angles have propagated different lengths inside the earth
so the amount of attenuation is different for them. In other
words, to be precise, the zenith angle dependence of $E_\nu^{cut}$
has to be taken into account.

\item  The energy threshold for detecting the neutrino also
depends on the direction. For the vertically propagating muon at
ICECUBE, $E_{th}^\mu$ is about 20 GeV; that is while, for the
muons propagating horizontally, $E_{th}^\mu$ is about 100 GeV
\cite{ICECUBE}.

\item  For the high energy muons with $E_\mu>1~$TeV, the muon ranges both in
the ice and rock exceed 1 km, (see Eq.~(\ref{icerange}) and
Ref.~\cite{range}). The depth of the ice  at the site of ICECUBE
is about 2810 m; so a considerable number of muons reaching the
ICECUBE would be produced in the rock beneath the ice where the
density is quite different.

\item As mentioned before, for neutrinos with energies higher than
100~{\rm TeV} the Earth is opaque. For very high energies,
$\nu_\tau$ can however be regenerated through $\nu_\tau \to \tau
\to \nu_\tau \to ... \to \nu_\tau$. As a result, $\nu_\tau$ and
$\bar{\nu}_\tau$ with $E_\nu \gg 100$~TeV can give a contribution
to the upward-going neutrino flux with $E<100$~TeV. The neutrino
flux at high energies is expected to be suppressed. Thus, such a
contribution is expected to be negligible \cite{tauneutrino}. This
assumption can in principle be tested by measuring the
downward-going shower-like events (which have not traversed the
Earth).
\end{itemize}
Throughout the present analysis, we use the approximate formulae
(\ref{muCC1},\ref{tauCCmu},\ref{shower},\ref{shower1},\ref{shower2}).
We determine how much the uncertainties in various inputs have to
be improved in order not to be an obstacle for  determination of
$\delta$ and $s_{13}$.  As we shall see, our conclusion is that
even without the above subtleties, the required precision  in
certain input parameters is so fine that seems beyond reach in the
foreseeable future. Taking into account the above uncertainties
not only will not change our conclusion but will  further confirm
it.


\section{The standard picture \label{standard}}

To evaluate $R$, several input parameters have to be known: i) the
energy spectrum of the incoming neutrinos; ii) the ratio of the
neutrino flux to the anti-neutrino flux; iii) the initial ratio
$w_e^0:w_\mu^0:w_\tau^0$. We rely on the predictions of the models
for such input parameters. Although the models differ in details,
they share some common features. From now on, we call these
features the ``standard picture.'' The features of the standard
picture are enumerated below.
\begin{itemize}
\item In the standard picture neutrinos are produced in the
following chain of processes. First, the energetic protons in jets
collide on $\gamma$ or on the background protons and produce
$\pi^+$ and $\pi^-$. Then, \ba \label{pion}
\begin{matrix} \pi^+&\to& \mu^+\nu_\mu ~ & ~~~~~~~~&\pi^- &\to
&\mu^- \bar{\nu}_\mu ~\cr \mu^+&\to & e^+ \bar{\nu}_\mu \nu_e &
~~~~~~~~ & \mu^-&\to & e^- {\nu}_\mu \bar{\nu}_e
\end{matrix}\ .\ea
Thus, $w_e^0:w_\mu^0:w_\tau^0=1:2:0$.
\item The energy spectra of the neutrinos follow power law
distributions:
\begin{equation} \label{power-law}
\frac{ dF_{\nu_\beta}}{dE_{\nu_\beta} }=\mathcal{N}_\beta
E_{\nu_\beta}^{-\alpha}\end{equation} where $\mathcal{N}_\beta$ is
the normalization factor for each neutrino and anti-neutrino
flavor. $\alpha$ is the spectral index. In the standard picture
where the initial protons are accelerated to high energies via
Fermi acceleration mechanism, the spectral index is expected to be
equal to 2 \cite{fermiacc}\ .

\item
Regardless of the relative amount of $\pi^+$ and $\pi^-$, we
expect $\mathcal{N}_{\nu_\mu}=\mathcal{N}_{\bar{\nu}_\mu}$.
However, $\mathcal{N}_{\bar{\nu}_e}/\mathcal{N}_{\nu_e}$ depends
on the initial composition of $\pi^+$ to $\pi^-$ which in turn
depends on the details of the model.
\end{itemize}

\section{Uncertainties and their impact on $\theta_{13}$ measurement \label{un-impact}}

Our knowledge of the sources of the cosmic neutrinos is quite
limited and mostly speculative. A myriad of known and un-known
effects can cause deviation of the initial flux from the standard
picture that was described in the previous section. In this
section, we compare the effect of a deviation from the standard
picture on $R$ with the effect of a nonzero $s_{13}$. Here, we
assume that the neutrino mass matrix conserves CP. A discussion of
CP-violation is given in sect.~\ref{degeneracy}.

Let us fix our convention for the mixing angles. Here, we use the
standard parametrization of PDG \cite{pdg} for the neutrino mixing
matrix with \cite{schwetz}
$$ 0\leq \theta_{13} <0.2< {\pi \over 2} \ \ \ \ {\rm and} \ \ \ \ 0\leq\delta<2\pi \ .$$
The sensitivity of $P_{\alpha \beta}$ on the phase $\delta$ is
through $\sin \theta_{13} \cos \delta$; so the CP-conserving cases
with $\delta=0$ and $\delta=\pi$ will have distinct effects. We
will consider both cases. Throughout this section we take
$w^0_e:w^0_\mu:w^0_\tau=1:2:0$.

In sect.~\ref{spectrum}, we discuss the effect of the variation of
the spectrum on ratio $R$. In sect.~\ref{antineutrino}, we discuss
the dependence of $R$ on
$\mathcal{N}_{\bar{\nu}_e}/\mathcal{N}_{\nu_e}$. Sect.~\ref{cross}
gives a brief discussion of the neutrino nucleon uncertainty and
its effects.

\subsection{Energy spectrum of incoming flux\label{spectrum}}

As mentioned before, in the standard picture, the neutrino flux
follows a power-law spectrum of form Eq.~(\ref{power-law}) with
$\alpha=2$. However, more careful considerations of the details of
the Fermi acceleration and the properties of the target particles
show that $\alpha$ can deviate from 2 and take any value in the
range (1,3) \cite{Melonialpha,Rachen,Learned}.

The energies of the muons and showers entering a neutrino
telescope can be measured. However, extracting the energy of the
incoming neutrinos that induce such events is not straightforward.
In the case of $\mu$-track events, the muon can lose a substantial
part of its energy before entering the detector. On the other
hand, limiting the analysis to the muons produced inside the
detector will reduce the statistics. In the case of the
shower-like events originating from the NC interaction of
neutrinos, the energy of the shower does not give the energy of
the initial neutrino because a part of the energy is carried away
by the final neutrino which escapes detection. Nevertheless, it is
shown in \cite{Beacom} that for $E_\nu^2dF_\nu/dE_\nu=0.25$ GeV
cm$^{-2}$ sr$^{-1}$ yr$^{-1}$ after one year of data-taking,
$\alpha$ can be determined with 10~\% uncertainty.

\begin{figure}[h!]
  \centering
  \includegraphics[bb=200 70 550 550,keepaspectratio=true,clip=true,angle=-90,scale=0.7]{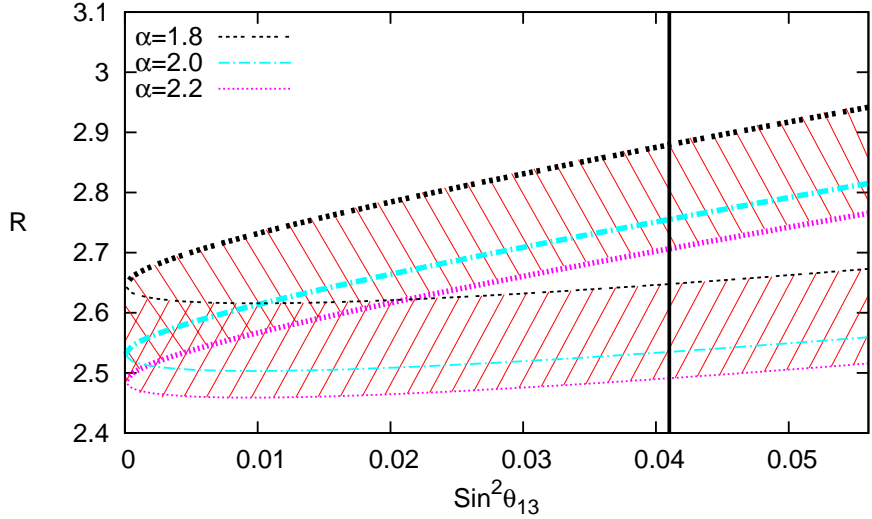}

  \caption{{\small The dependence of $R$ on $\sin^2\theta_{13}$
  for different values of
  the  spectral index, $\alpha$. The thicker lines correspond to
  $\delta=\pi$ and the thinner ones correspond to $\delta=0$.
 We have used the central values for the
  neutrino-nucleon cross section \cite{Gandhi} and have set $\mathcal{N}_{\bar{\nu}_e}/\mathcal{N}_{\nu_e}=0.5$ and
  $(\mathcal{N}_{\bar{\nu}_\mu}+\mathcal{N}_{\nu_\mu})/(\mathcal{N}_{\bar{\nu}_e}+\mathcal{N}_{\nu_e})=2$.
  The input for $\theta_{12}$ and $\theta_{23}$ are set equal to the best
  fit in \cite{schwetz}.
  The vertical line at 0.041 shows the present bound at 3$\sigma$ \cite{schwetz}.} }
  \label{fig1,2}
\end{figure}

Fig.~\ref{fig1,2} shows $R$ versus $\sin^2 \theta_{13}$ for $\cos
\delta=\pm 1$ and various values of $\alpha$. As seen from the
figure when $\delta=0$, the sensitivity of $R$ to $s_{13}^2$ is
very mild and less than 2~\%. That is while for $\cos \delta=-1$,
the sensitivity to $s_{13}^2$ is about 10~\%. The disparity
between $\cos \delta=+1$ and $\cos \delta=-1$ means that for
$s_{13}^2\sim 0.04$, the contributions from $s_{13}\cos \delta$
and $s_{13}^2$ are comparable. In fact, expanding $R$ in powers of
$s_{13}$ confirms this claim: \be R\simeq r_1+r_2 s_{13}\cos
\delta +r_3 s_{13}^2\cos^2 \delta, \label{expansion-of-R} \ee
where  for the central curve with $\alpha=2.0$, $r_1\simeq 2.55$,
$r_2\simeq -0.66$ and $r_3\simeq 2.65$. If $\theta_{23}$ deviates
from $\pi/4$, in addition to the $s_{13}^2\cos^2 \delta$ term, a
term proportional to $s_{13}^2$ has to be added to
Eq.~(\ref{expansion-of-R}).

As seen from the figure, even for $\cos \delta=-1$, the
sensitivity to $s_{13}^2$ can be obscured by the 10~\% uncertainty
in $\alpha$. However, for $s_{13}^2>0.02$, the bands between
$\alpha=2.2$ and $1.8$ for $\cos \delta=1$ and $\cos \delta=-1$
have no overlap. This means that for $s_{13}^2>0.02$, 10~\%
precision in $\alpha$ is enough to distinguish $\cos \delta=1$
from $\cos \delta=-1$.

Notice that the curve with $\alpha=2$ is closer to that with
$\alpha=2.2$ than that with $\alpha=1.8$. This means that the
effect of the uncertainty decreases by increasing the value of
$\alpha$.

\begin{figure}[h!]
  \centering
  \includegraphics[bb=250 70 550 550,keepaspectratio=true,clip=true,angle=-90,scale=0.8]{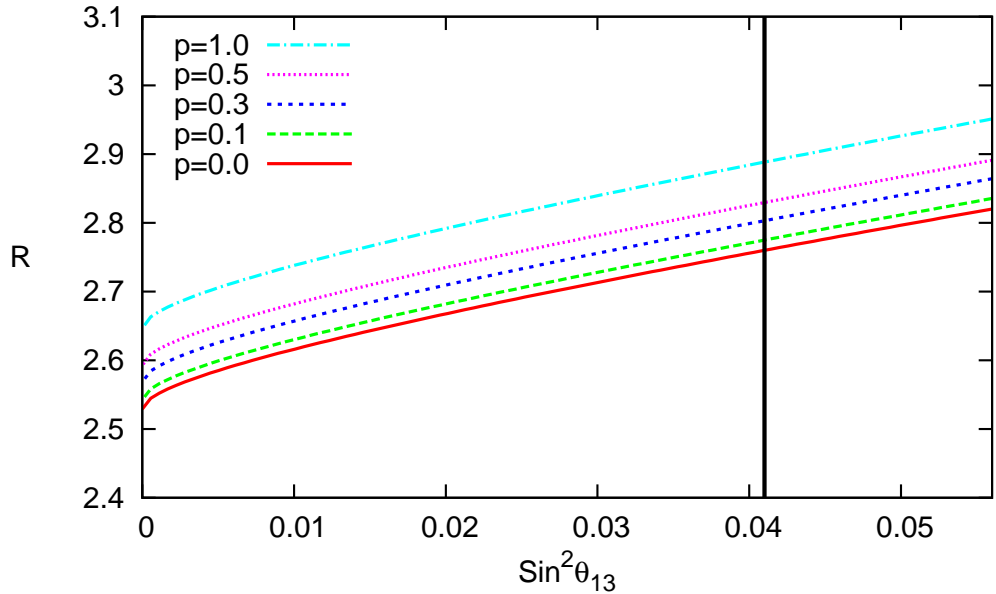}

  \caption{{\small The dependence of $R$ on $\sin^2\theta_{13}$
  for different values of the $p$ parameter defined in
  Eq.~(\ref{fist-second}). We have set $\delta=\pi$. The rest of the
  input parameters
  are the same as in Fig.~\ref{fig1,2}.   The vertical line at 0.041 shows the present bound at 3$\sigma$ \cite{schwetz}.} }
  \label{fig3}
\end{figure}

Considering the unknown nature of the production mechanism, it is
not dismissed that the energy spectrum of neutrinos does not
follow a simple power-law form. For example the spectrum can be a
sum of two power-law functions each originating from a separate
mechanism ($i.e., a E^{-\alpha_1}+b E^{-\alpha_2}$). We study such
a possibility in Fig.~\ref{fig3}, where we have taken the shape of
the spectrum to be of the form: \be \label{fist-second} E^{-2}
+p\left(\frac{E^{-1}}{100~{\rm TeV}}\right)  \ , \ee where $p$ is
a dimensionless parameter that determines the  magnitude of the
second term.
 Both terms
can originate from the Fermi acceleration mechanism
\cite{Gaisser}.  Notice that the second term is subdominant. The
curves from up to down respectively correspond to $p=1$, 0.5, 0.3,
0.1 and 0. As seen from the figure, the uncertainty on $p$
obscures the extraction of the mixing angle $\theta_{13}$.

\subsection{Uncertainty in
$\mathcal{N}_{\bar{\nu}_e}/\mathcal{N}_{\nu_e}$
\label{antineutrino}}

\begin{figure}[h!]
  \centering
  \includegraphics[bb=250 70 550 550,keepaspectratio=true,clip=true,angle=-90,scale=0.8]{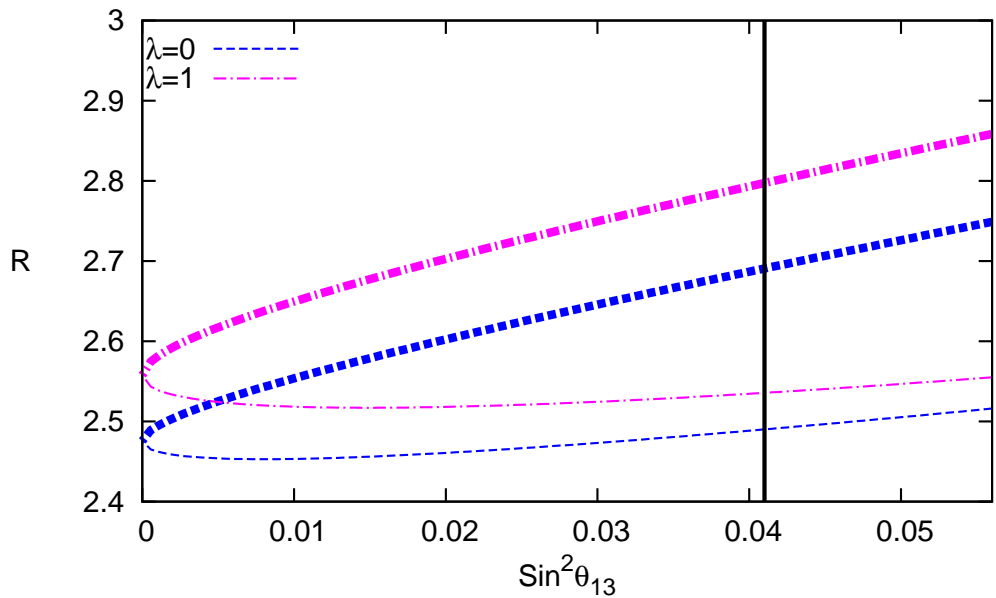}

  \caption{{\small The dependence of $R$ on $\sin^2\theta_{13}$
  for different values of the parameters
  $\lambda\equiv\mathcal{N}_{\bar{\nu}_e}/\mathcal{N}_{\nu_e}$.
    The thicker lines correspond to
  $\delta=\pi$ and the thinner ones correspond to $\delta=0$.
  The spectral index has been set equal to $2$. The rest of the input parameters are the
  same as in Fig.~(\ref{fig1,2}).} }
  \label{fig4}
\end{figure}

Since the source is made of matter rather than anti-matter, we in
general expect $\pi^+$ to dominate over $\pi^-$ and therefore $ 0<
\mathcal{N}_{\bar{\nu}_e}/\mathcal{N}_{\nu_e}<1$. There is not any
established or proposed method for determining
$\mathcal{N}_{\bar{\nu}_e}/\mathcal{N}_{\nu_e}$ in the neutrino
telescopes in the energy interval $(100~{\rm GeV},100~{\rm TeV})$.
As a result, this ratio appears as a source of uncertainty in
determination of $R$. Curves in the Fig.~\ref{fig4} show the
dependence of $R$ on $\sin^2\theta_{13}$ for two extreme values
$\lambda\equiv \mathcal{N}_{\bar{\nu}_e}/\mathcal{N}_{\nu_e}=0$
and $\lambda=1$. For $\delta=\pi$, the variation of
$\mathcal{N}_{\bar{\nu}_e}/\mathcal{N}_{\nu_e}$ in the interval
$[0,1]$ causes a change in $R$ of about 5~\% which can obscure the
determination of $s_{13}$. Notice that for $s_{13}^2>0.005$ the
bands between $\lambda=1$ and $\lambda=0$ for $\delta=0$ and
$\delta=\pi$ are separate, so the uncertainty in $\lambda$ will
not cause a problem for discriminating between $\cos  \delta=\pm
1$.

\subsection{Uncertainties in cross sections\label{cross}}

To calculate the cross section, information on the Parton
Distribution Functions (PDFs) of the nucleon is needed. The center
of mass energy of a system composed of a neutrino with energy
$E_\nu\sim 100$ TeV incident on a proton at rest is  $(2 E_\nu
m_p)^{1/2}=450$~GeV. The center of mass energy of the $e-p$ HERA
collider is about 320 GeV. Thus, to calculate $\sigma_{\nu N}$ in
the energy range relevant for this study ({\it i.e.,} $100$ GeV
$\lesssim E_\nu \lesssim 100$ TeV) the results of the HERA
experiment can be employed. The current uncertainty  on the PDFs
is about 3~\%  \cite{cooper-sarkar}. LHC can further improve the
precision of the PDFs.

{The uncertainty in the cross section of all types of neutrinos
(each flavor of neutrino and anti-neutrino) originates from the
same uncertainties in the PDFs. As a result, the resultant
uncertainty in the numerator and denominator of $R$ cancel each
other so the certainty in the cross section will not be a limiting
factor for determining $\theta_{13}$ (or $\delta$) from the cosmic
neutrino flavor composition.}

\section{ Determination of $\delta$ \label{degeneracy}}

In the literature it is suggested to derive the Dirac CP-violating
phase ($\delta$) by studying the flavor composition of the cosmic
neutrinos \cite{parameters,theta23,Farzan1}. Considering the
expenses and challenges before measuring this phase through the
more conventional proposals ({\it i.e.,} neutrino factory or
superbeam methods), it is worth giving this possibility a thorough
consideration. However, in the literature the flavor
identification power of the detector has not been realistically
treated. To be specific, the quantities that have been previously
analyzed in the context of deriving $\delta$  are ratios such as
$R'\equiv F_{\nu_\mu}/(F_{\nu_e}+F_{\nu_\tau})$
\cite{parameters,theta23,Farzan1} which cannot be directly derived
at neutrino telescopes.
In this section, we assess the possibility of measuring $\delta$
considering realistic flavor identification power of neutrino
telescopes ({\it i.e.,} studying ratio $R$) and taking into
account various sources of uncertainty for the first time.

By the time a statistically significant number of cosmic neutrino
events is collected, we expect noticeable improvement in
determination of the input parameters. In particular, in the case
that the parameters are in favorable range, we expect progress in
the following measurements:

\begin{itemize}
\item
If $\sin^2\theta_{13}$ is close to the present bound, the
forthcoming experiments can measure its value with a precision of
$\Delta \sin^2 \theta_{13}$:
$$\sin^2 \theta_{13}=\sin^2 \bar{\theta}_{13}(1 \pm \Delta\! \sin^2 \theta_{13}/\sin^2 \bar{\theta}_{13}).$$
In fact, for relatively large values of $\sin^2\theta_{13}$, the
uncertainty $\Delta \! \sin^2 \theta_{13}/\sin^2
\bar{\theta}_{13}$ can be reduced to as small as 5 \%
\cite{Kozlov,Ardellier}.
\item
In case that statistically significant number of cosmic neutrinos
are collected, $R$ can be measured with an uncertainty of $\Delta
R$:
$$R=\bar{R}(1\pm \Delta R/\bar{R}).$$
As shown in \cite{Beacom}, a precision of $\Delta R/\bar{R}\simeq
7\%$ can be obtained provided that the number of events exceeds
$\sim$ 300. This would be achieved with a neutrino flux of
$E_\nu^2dF_\nu/dE_\nu=0.25$ GeV cm$^{-2}$ sr$^{-1}$ yr$^{-1}$
after a couple of years of data-taking.

\item
The forthcoming long-baseline \cite{nova} and reactor neutrino
\cite{kamland} experiments can respectively measure the solar and
atmospheric mixing angles by a precision of  $\sim$~6~\%. That is
$$ \sin^2\theta_{12}=\sin^2\bar{\theta}_{12}(1\pm 6\%),$$
$$ \sin^2\theta_{23}=\sin^2\bar{\theta}_{23}(1\pm 6\%).$$ The
present best-fit values are $\sin^2\bar{\theta}_{12}=0.32$ and
$\sin^2\bar{\theta}_{23}=0.5$.

\end{itemize}

Remember that the ratio
$\mathcal{N}_{\bar{\nu}_e}/\mathcal{N}_{\nu_e}$ cannot be
measured. Since the initial jets creating the charged pions (and
subsequently the neutrinos) are mainly made of protons rather than
anti-protons, we expect
$\mathcal{N}_{\bar{\nu}_e}/\mathcal{N}_{\nu_e}\leq 1.$ Considering
all these uncertainties, the question is whether it will be
possible to extract the value of $\delta$.

\begin{figure}[h!]
  \centering
  \includegraphics[bb=250 70 600 570,keepaspectratio=true,clip=true,angle=-90,scale=0.7]{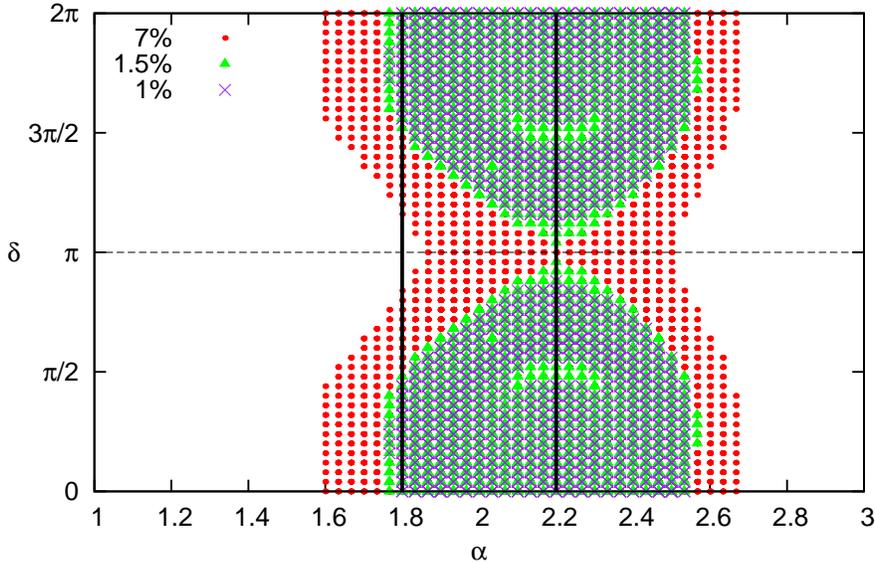}

  \caption{{\small Points in the $(\alpha,\delta)$ space
consistent with $R=2.53 \pm \Delta R$. True values of the
$(\alpha,\delta)$ pair are $(2,\pi/2)$. Points displayed by dots,
 triangles and crosses respectively correspond to
  $ \Delta R/\bar{R}=7\%$, $ \Delta
 R/\bar{R}=1.5\%$ and $ \Delta
 R/\bar{R}=1\%$. To draw this figure we have varied
$\sin^2\theta_{13} \in (0.028,0.032)$,
  $\sin^2\theta_{12} \in (0.30,0.34)$, $\sin^2\theta_{23} \in (0.47,0.53)$
   and $\mathcal{N}_{\bar{\nu}_e}/\mathcal{N}_{{\nu}_e}\in(0,1)$.} }
  \label{delta}
\end{figure}

\begin{figure}[h!]
  \begin{center}
  \centerline{\vspace{0.5cm}}
 \centerline{\includegraphics[bb=260 75 570
 500,keepaspectratio=true,clip=true,angle=-90,scale=0.58]{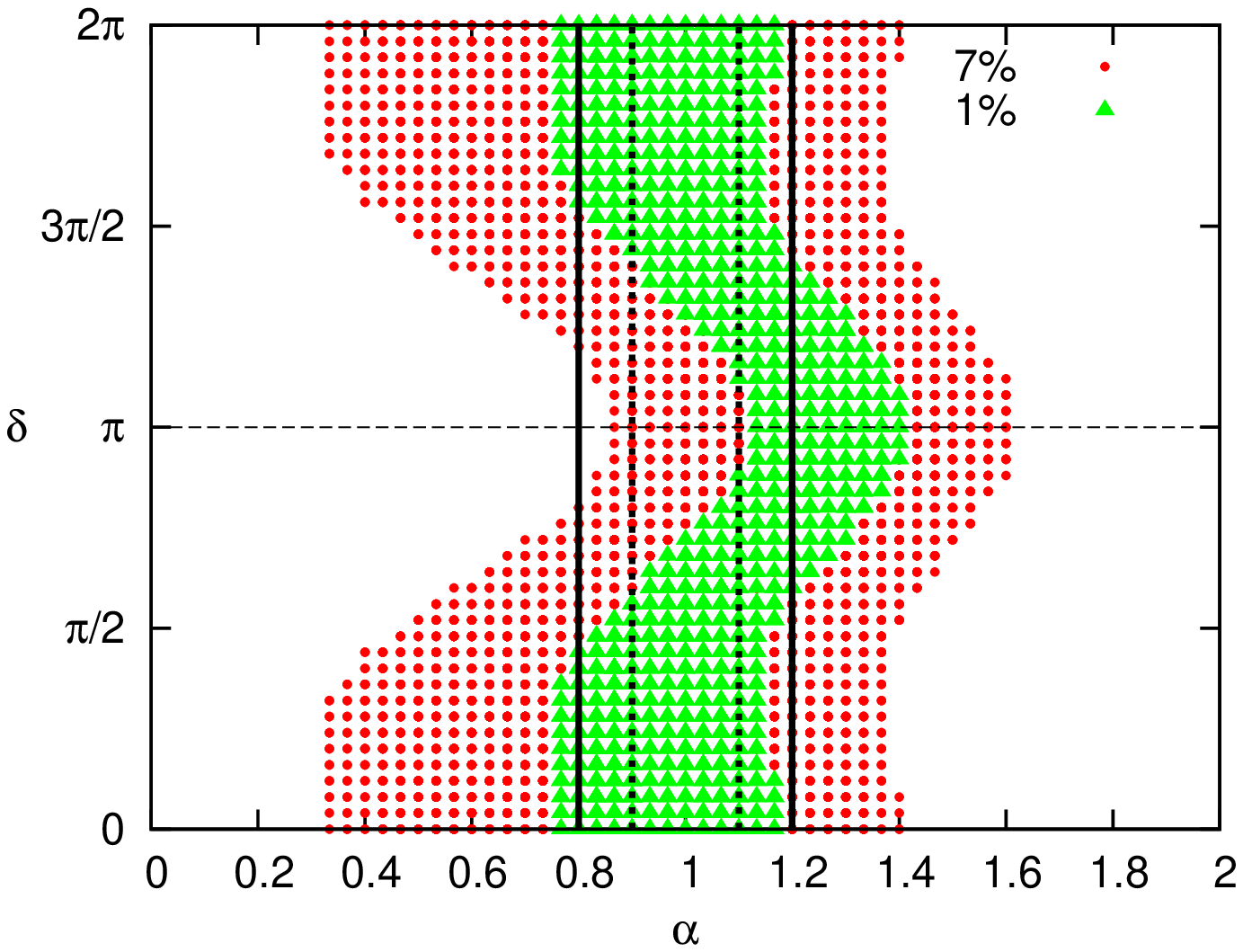}\includegraphics[bb=260 70 570
 500,keepaspectratio=true,clip=true,angle=-90,scale=0.58]{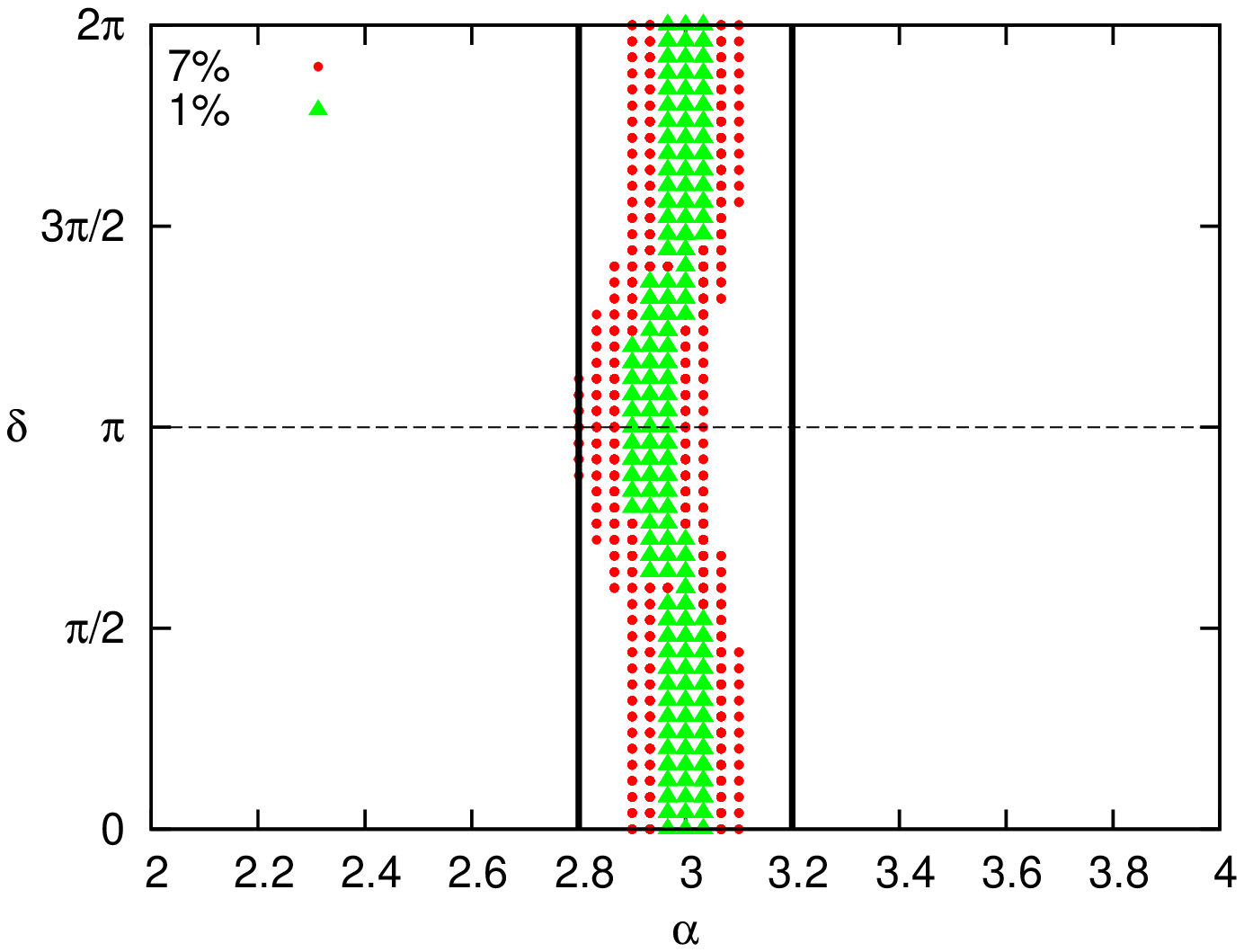}}
 \centerline{\hspace{0.8cm}(a)\hspace{8.5cm}(b)}
 \centerline{\vspace{-1.5cm}}
 \end{center}
 \caption{{\small The same as Fig.~\ref{delta} except
  that  in figure (a), we have taken $\bar{R}=3.30$ (corresponding to
 $\alpha=1$) and in  figure (b), we have taken $\bar{R}=3.59$ (corresponding to
 $\alpha=3$).
 Points displayed by dots and
 triangles  respectively correspond to
  $ \Delta R/\bar{R}=7\%$ and $ \Delta
 R/\bar{R}=1\%$.} }
  \label{alpha1va3}
\end{figure}
\begin{figure}[h!]
  \centering
  \includegraphics[bb=250 70 600 570,keepaspectratio=true,clip=true,angle=-90,scale=0.7]{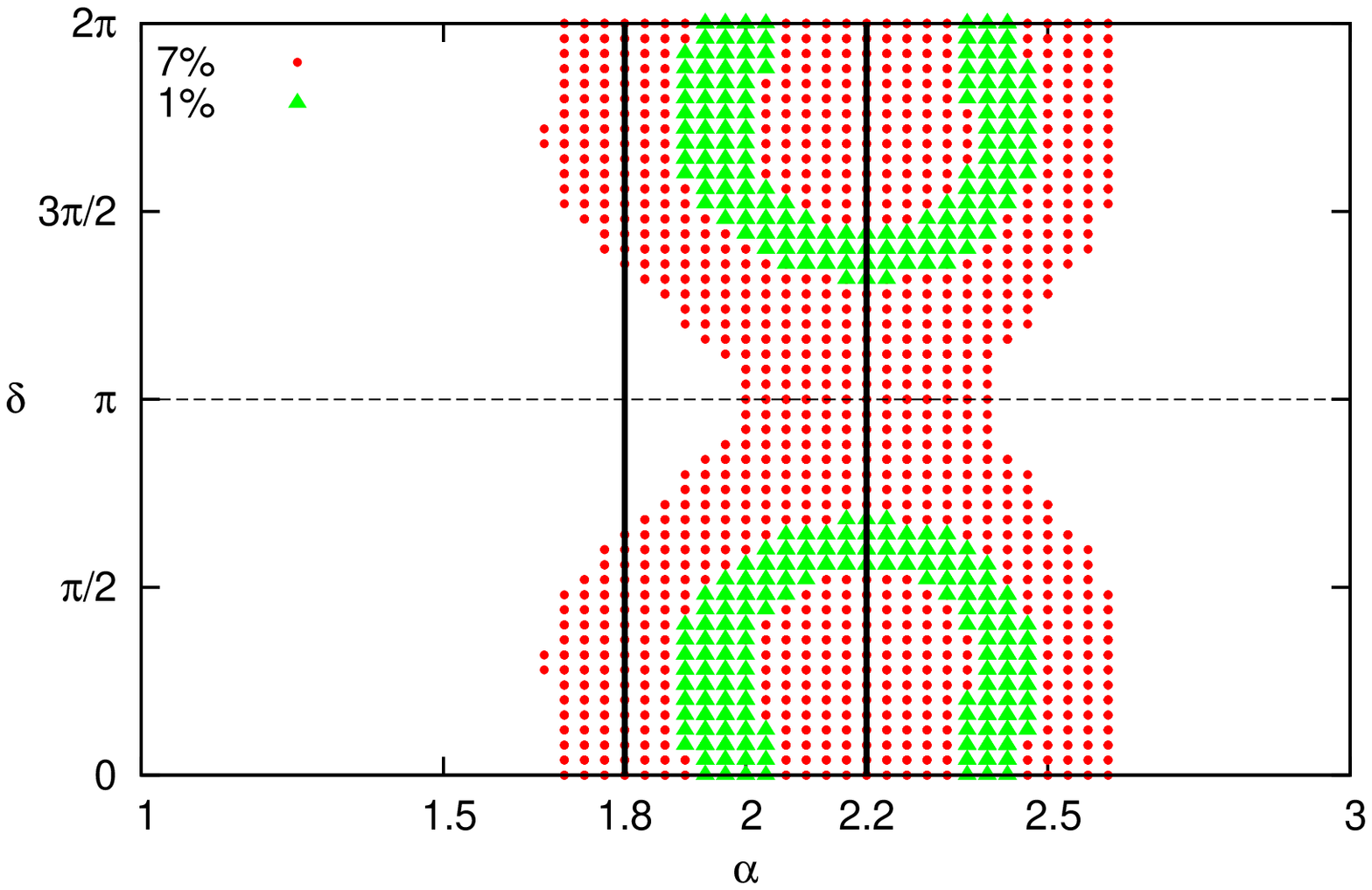}
  \caption{{\small Points in the $(\alpha,\delta)$ space
consistent with $R=2.53 \pm \Delta R$.  As in Fig.~\ref{delta},
the true values of the $(\alpha,\delta)$ pair are $(2,\pi/2)$.
Points displayed by dots
 and triangles respectively correspond to
  $ \Delta R/\bar{R}=7\%$ and $ \Delta
 R/\bar{R}=1\%$. To draw this figure we have fixed
$\sin^2\theta_{13}=0.0301$,
  $\sin^2\theta_{12}=0.32$,
  $\mathcal{N}_{\bar{\nu}_e}/\mathcal{N}_{{\nu}_e}=0.5$ and
  varied
  $\sin^2\theta_{23}$ in $0.5(1\pm 1\%)$.
   } }
  \label{revised}
\end{figure}

Fig.~\ref{delta} addresses this question. Drawing the plot, we
have assumed that $\bar{R}$ will be found to have a typical value
of 2.53 with an uncertainty of $\Delta R/\bar{R}$. This value of
$\bar{R}$ can be obtained by taking maximal CP-violation
$(\delta=\pi/2)$, $\sin^2\theta_{13}=0.03$,
$w_e^0:w_\mu^0:w_\tau^0=1:2:0$, $\alpha=2$ and
$\mathcal{N}_{\bar{\nu}_e}/\mathcal{N}_{{\nu}_e}=0.5\ $. We have
looked for solutions in the $\delta-\alpha$ plane for which
$R=2.53(1\pm\Delta R/\bar{R})$, varying the rest of the relevant
parameters in the ranges indicated in the caption of
Fig.~\ref{delta}. The regions covered with dots, little triangles
and crosses respectively correspond to 7\%, 1.5\% and 1\%
precision in the measurement of $R$.  Notice that the figure is
symmetric under $\delta \to 2\pi -\delta$. The symmetry originates
from the fact that the dependence of $R$ on $\delta$ is through
$\cos \delta$. As mentioned earlier, $\alpha$ can be independently
determined by the measurement of the energy spectrum with about
10\% precision. (For $\alpha=2$, the direct measurement of the
energy spectrum can restrict the value of $\alpha$ to the region
between the vertical lines at $\alpha=1.8$ and 2.2.) As seen from
the figure, with $\Delta R/\bar{R}=7 \%$, $\delta$ cannot be
constrained. In fact, any point between the vertical lines can be
a solution. The figure shows that reducing $\Delta R/\bar{R}$ to 1
\% (but keeping the rest of the uncertainties as before), some
parts of the solutions can be excluded. In particular, the region
around $\delta=\pi$ will not be a solution anymore. Notice that
along with $\delta =\pi/2$, $\delta=0$ is also a solution. This
means that despite maximal CP-violation, the CP-violation cannot
still be established. We examined the robustness of this result.
We found that reducing the uncertainties in the mixing angles
(even in $\theta_{13}$) will not noticeably change the overall
conclusion. However, the sensitivity to $\Delta{R}/\bar{R}$ seems
to be high. Notice that with a precision of
$\Delta{R}/\bar{R}=1.5\%$, there are some regions of solutions
(covered by the triangles) that can be excluded if the uncertainty
is reduced to 1\% ({\it i.e.,} they are not covered with crosses).
As we will see below, the sensitivity to power index is also high.

Drawing Figs.~(\ref{alpha1va3}-a,\ref{alpha1va3}-b), we have
respectively taken $\bar{R}=3.3$ (corresponding to $\alpha=1$) and
$\bar{R}=3.59$ (corresponding to $\alpha=3$). The rest of the
input is the same as Fig.~(\ref{delta}). From
Fig.~(\ref{alpha1va3}-b) we observe that for $\alpha=3$, the
measurement of $R$ with 7\% uncertainty determines $\alpha$ with
better than 6 \% precision which will probably be  more accurate
than the direct determination of $\alpha$ from the energy spectrum
measurement. Notice that when $\alpha=1$ or 3, even a precision of
$\Delta R/\bar{R} =1 \%$ will not be enough to constrain $\delta$.
However, from Fig.~(\ref{alpha1va3}-a) we observe that in case of
$\alpha=1$ and $\delta=\pi/2$, if the error in direct measurement
of $\alpha$ is reduced to $10 \%$ ({\it i.e.}, if $\alpha$ is
constrained to the region between the dashed vertical lines) and
if $R$ is measured with 1\% precision, one can exclude solutions
around $\delta=\pi$ but CP-violation cannot still be established.

As emphasized in \cite{theta23}, the sensitivity of $R$ to the
variation of  $\theta_{23}$ is significant. In fact, the present
uncertainty ({\it i.e.,} $\sin^2\theta_{23}=0.50^{+0.18}_{-0.16}$
at $3\sigma$ c.l. \cite{schwetz}) leads to a sizeable uncertainty
of $\sim 20$ \% in $R$.
As mentioned earlier the forthcoming experiments \cite{kamland}
can improve the precision of $\sin^2 \theta_{23}$ to 6 \%. A
variation of 6\% in $\sin^2\theta_{23}$ leads to a $\sim 4$ \%
change in $R$ which is comparable to the effect of $\cos \delta$
for $\sin^2 \theta_{13}=0.03$. To pinpoint the effect of the
uncertainty of $\theta_{23}$ on $R$, we have presented
Fig.~\ref{revised}. The input of Fig.~\ref{revised} is similar to
Fig.~\ref{delta} but to study the effect of $\theta_{23}$, we have
fixed $\mathcal{N}_{\bar{\nu}_e}/\mathcal{N}_{{\nu}_e}$,
$\theta_{12}$ and $\theta_{13}$ to their central values. Comparing
Figs.~\ref{delta} and \ref{revised}, we find that improving the
precision of $\sin^2 \theta_{23}$ from 6\% to 1\% can help us to
remove a substantial part of the spurious solutions. Especially
for $\Delta R/ \bar{R}=1\%$ solutions with $3\pi/4<\delta<5\pi/4$
can be removed by reducing the uncertainty of $\sin^2\theta_{23}$
to 1~\%.

\section{Initial flavor composition \label{flavor-composition}}

In the previous sections, in accordance with the ``standard
picture'', we had assumed that $w_e^0:w_\mu^0:w_\tau^0=1:2:0$.
Various mechanisms can intervene to cause a deviation from this
simple prediction. For example, in the so-called stopped muon
scenario, the muon production takes place inside a high magnetic
field so the muons come to rest before decay. As a result, the
neutrinos produced from the decay of the muon would be below the
detection energy threshold. This effectively leads to
$w_e^0:w_\mu^0:w_\tau^0=0:1:0$ \cite{Rachen}. On the other hand, a
contribution from neutron decay, $n\to \bar{\nu}_e p e^-$, would
increase $w_e^0/w_\mu^0$. Even in the standard picture, along with
the production of $\pi^\pm$, charged Kaons can  also be produced
in the jets. Like the case of the charged pions, the main decay
mode of charged Kaon is $K\to \mu \nu_\mu$, so neglecting the
subdominant modes Br($K\to \pi^0 \mu \nu_\mu)=3.3 \%$ and Br($K\to
\pi^0 e \nu_e)=5 \%$ \cite{pdg}, we would naively expect that
flavor composition of the neutrinos from the Kaon chain also
follow the standard picture (i.e., $w_e^0:w_\mu^0:w_\tau^0\simeq
1:2:0$) but there is a subtlety here. In the chain process, $\pi
\to \mu \nu_\mu$, $\mu \to \nu_e \nu_\mu e$, the average energy of
each of the produced neutrinos  in the rest frame of the pion is
about $m_\pi/4$. That is while in the case of Kaon decay, the
average energies of the neutrinos are different: The energy of
$\nu_\mu$ produced directly from the Kaon decay in the Kaon rest
frame is $(m_K^2-m_\mu^2)/(2m_K)\simeq 236$~ MeV; that is while,
the average energies of the neutrinos from the secondary muon are
$[(m_K^2+m_\mu^2)/(2 m_K)]/3\simeq 86$~MeV. As a result, limiting
the detection to an energy range (e.g., (100 GeV,100TeV)) the
ratio $w_e^0:w_\mu^0~
(i.e.,~\int_{E_{th}}^{E_{cut}}(dF_{\nu_e}^0/dE
+dF_{\bar{\nu}_e}^0/dE)dE:\int_{E_{th}}^{E_{cut}}(dF_{\nu_\mu}^0/dE
+dF_{\bar{\nu}_\mu}^0/dE)dE)$ will deviate from 1:2. In fact, the
ratio $w_e^0:w_\mu^0$ would depend on the energy spectrum of the
initial charged Kaon.

The mechanisms that we pointed out above are all processes that
expected to exist and play at least a subdominant role within the
framework of the mainstream models. None of these mechanisms
creates $\tau$ neutrino or anti-neutrino at the sources ({\it
i.e.,} they all yield $w_e^0:w_\mu^0: w_\tau^0=w^0_e:w^0_\mu:0$).
However, more exotic mechanisms can be at work to create
$\nu_\tau$ or $\bar\nu_\tau$ at source: In principle, the
collision of $pp$ or $p \gamma$ at the source can create $D$ meson
whose decay can produce $\nu_\tau$ and $\bar{\nu}_\tau$. According
to \cite{reno}, the contribution of the $D$ meson to the neutrino
flux becomes important only for $E_\nu>10^5$ TeV. For lower values
of energy in which we are interested in, the contribution is about
three orders of magnitude below the present bound. To our best
knowledge, within the Standard Model (SM) of particles, in the
energy range 100~GeV-100~TeV, $w_\tau^0$ remains much smaller than
$w_\mu^0$ and $w_e^0$. However, if neutrinos have some yet
unexplored properties beyond SM, tau neutrinos or anti-neutrinos
can be produced at the source. For example, suppose neutrinos
posses tiny transition magnetic moments of form \be \mu_{\tau e}
F_{\alpha \beta} \nu_e^T C \sigma^{\alpha \beta}\nu_\tau \ \ \
{\rm and/or} \ \ \ \mu_{\tau \mu} F_{\alpha \beta} \nu_\mu^T C
\sigma^{\alpha \beta}\nu_\tau\ .\ee There are strong bounds on the
transition moments, $|\mu_{\alpha \beta}|$, from different
terrestrial experiments and astrophysical observations
\cite{pdg,Raffelt}: \be \mu_{\alpha \beta}<3\times 10^{-12} \mu_B,
\label{moment-bound} \ee  where $\mu_B$ is the Bohr magneton. The
magnetic field inside the source can be so large that even a
transition moment as tiny as $10^{-13}\mu_B$ can lead to a
sizeable production of tau neutrino and anti-neutrino at the
source \cite{Athar}. In fact according to the models
\cite{magnetic-field}, it is possible to have \be
\label{hadd}\left(\frac{B}{10^9~{\text
{Gauss}}}\right)\left(\frac{L}{10^8~
{\text{cm}}}\right)\left(\frac{\mu_{\alpha
\beta}}{10^{-13}~\mu_B}\right) \gtrsim 1 \ee where $B$ is the
magnetic field and $L$ is the linear size of the volume in which
the magnetic field is as large as $B$. If condition (\ref{hadd})
is fulfilled, the flavor composition at source will considerably
deviate from the standard $1:2:0$. Let us normalize the initial
flavor ratio such that $w_\mu^0=1$:
$$w_e^0:w_\mu^0: w_\tau^0=w^0_e:1:w^0_\tau\ .$$
In this section, we discuss whether by merely studying $R$ and
without theoretical prejudice, one can extract $w_e^0$ and
$w_\tau^0$. Actually, the possibility of deriving information on
$w_e^0$ and $w_\tau^0$ from the measurement of the flavor ratio at
the detector has been discussed in the literature
\cite{flavorratio}. Like the case of measurement of neutrino
parameters, the flavor identification power of the neutrino
telescopes has not been realistically treated in the previous
studies. Here we investigate this possibility considering
realistic flavor identification power of neutrino telescopes ({\it
i.e.,} studying ratio $R$) and taking into account various sources
of uncertainty for the first time.

There is a subtle point here. If within the lifetime of the
ICECUBE (or a more advanced neutrino telescope) neutrino flux from
a GRB in a close-by galaxy (a $\sim 5$~Mpc far away galaxy) is
detected, the statistics will be high enough to extract
information on the flavor composition of the flux from this
individual source. For sources located at cosmological distances
($\gtrsim 100$~Mpc), the best can be done is to combine the
information from different sources. Each source may be different
and emit neutrino flux with a different flavor composition. From
the ``average'' $R$, only ``average'' values of $w_e^0$ and
$w_\tau^0$ over these sources can be derived.

\begin{figure}[h!]
  \begin{center}
 \centerline{\includegraphics[bb=260 70 570
 500,keepaspectratio=true,clip=true,angle=-90,scale=0.58]{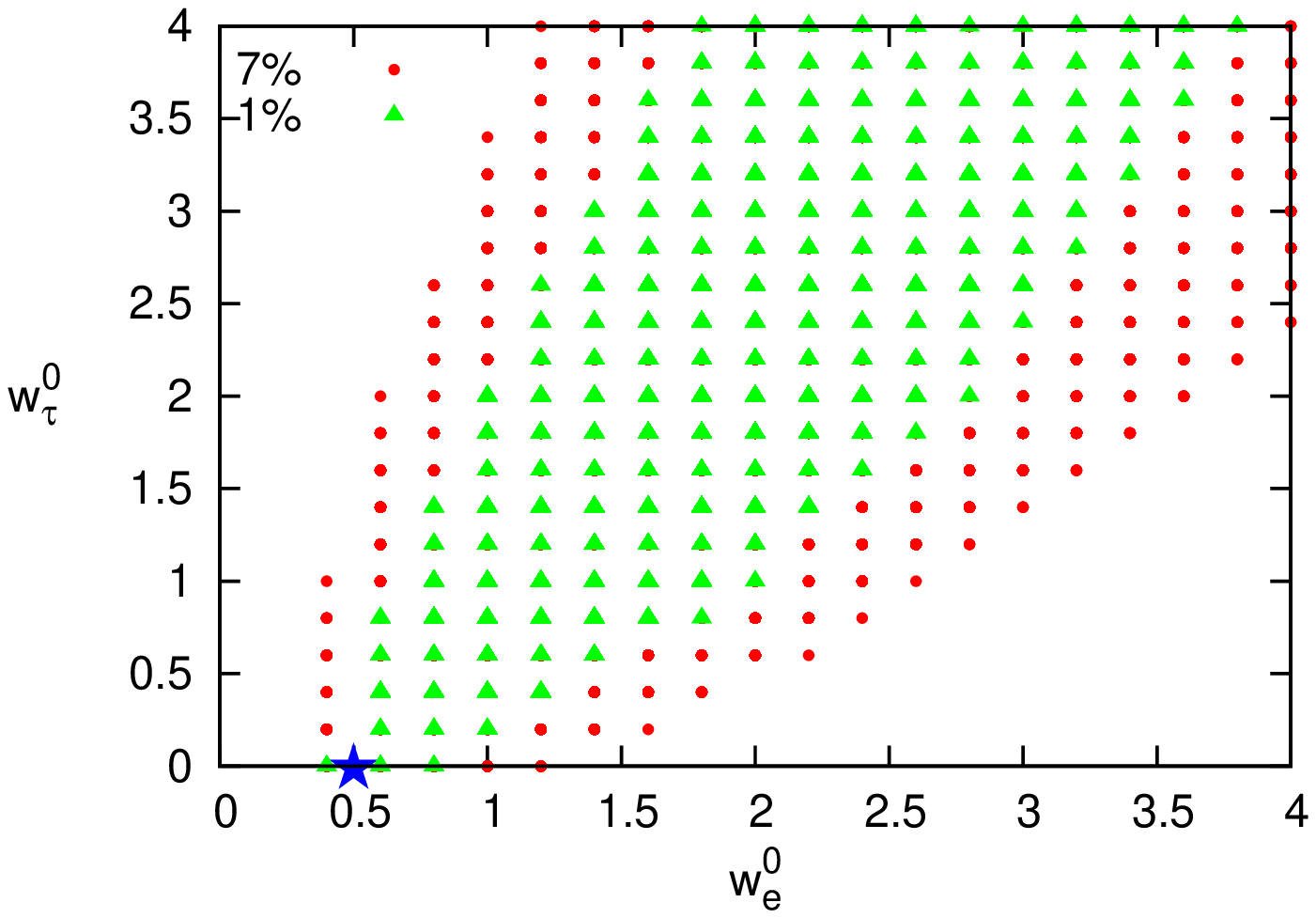}\includegraphics[bb=260 70 570
 500,keepaspectratio=true,clip=true,angle=-90,scale=0.58]{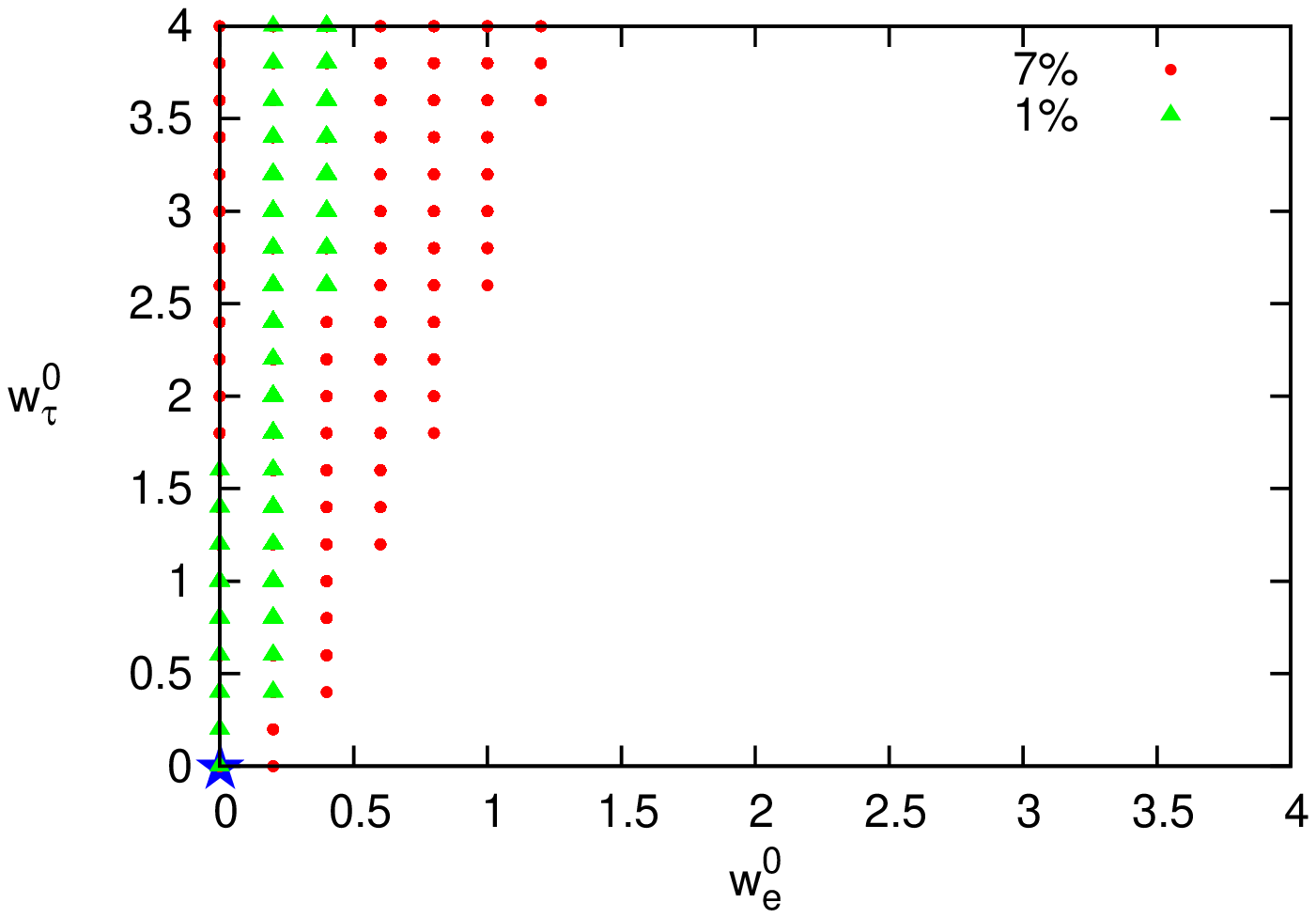}}
 \centerline{\hspace{0.8cm}(a)\hspace{8.5cm}(b)}
 \centerline{\vspace{-1.5cm}}
 \end{center}
 \caption{{\small Points in the $(w^0_e,w^0_\tau)$ plane consistent
with $R=\bar{R} \pm \Delta R$. The ratios are normalized such that
$w_\mu^0=1$. The true values of $(w^0_e,w^0_\tau)$ are denoted by
$\bigstar$.
 Points displayed by dots and
 triangles respectively correspond to $ \Delta R/\bar{R}=7\%$ and $ \Delta
 R/\bar{R}=1\%$. In drawing this figure we have varied
$\sin^2\theta_{13} \in (0,0.003)$,
  $\delta \in (0,2\pi)$, $\alpha \in (1.8,2.2)$,
   and $\mathcal{N}_{\bar{\nu}_e}/\mathcal{N}_{{\nu}_e}\in(0,1)$.
Drawing Fig.~(a), we have taken $\bar{R}=2.53$ which corresponds
to the standard picture with $w^0_e=1/2$ and $w^0_\tau=0$. In case
Fig.~(b), we have set $\bar{R}=3.2$ which corresponds to the
stopped muon
 scenario with $w^0_e=w^0_\tau=0$.
 } }
  \label{ab120}
\end{figure}

Taking into account the relevant uncertainties in the input
parameters, we look for values of $w_e^0:w_\mu^0:w_\tau^0$ that
are consistent with $R=\bar{R} \pm \Delta R$. To perform this
analysis, we take $\theta_{13}=0$. For any other value of
$\theta_{13}$, the same analysis can be repeated. The results are
robust against the variation of $\theta_{13}$ within the present
bound. If $\theta_{13}=0$, by the time that enough cosmic
neutrinos are collected, the Daya-Bay \cite{dayabay} and
Double-Chooz \cite{Ardellier} experiments can set the bound
$\sin^2 \theta_{13}<0.003$. We vary $\sin^2\theta_{13}$ between
zero and 0.003. In this case, there is no hope of measuring
$\delta$ so we allow $\delta$ to vary between 0 and $2\pi$. We
take the energy spectrum to be of form $E^{-2}$ and assume that
its power-law behavior will be established and the spectral index
will be measured with 10 \% precision. Again, we vary
$\mathcal{N}_{\bar{\nu}_e}/\mathcal{N}_{{\nu}_e} $ within [0,1].

\begin{figure}[h!]
  \centering
  \includegraphics[bb=260 70 550 520,keepaspectratio=true,clip=true,angle=-90,scale=0.7]{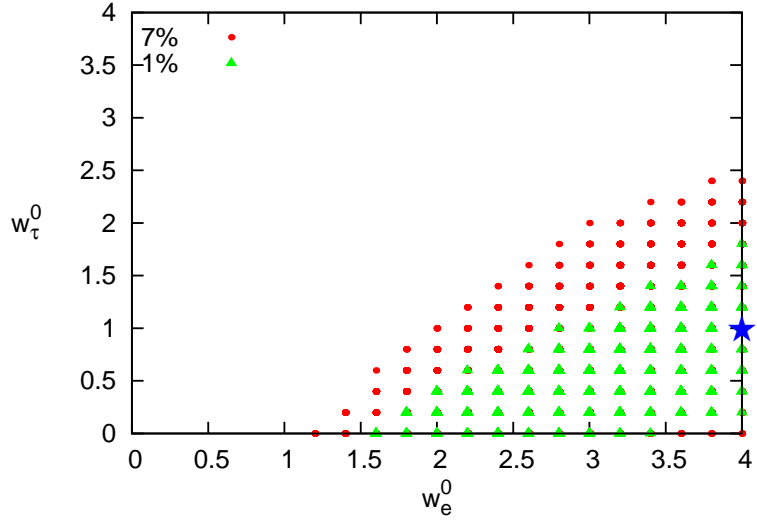}
 \caption{{\small The same as Fig.~\ref{ab120} except that $\bar{R}$ is set equal to 1.96, which corresponds to
 $w_e^0:w_\mu^0:w_\tau^0=4:1:1$.
 } }
  \label{ab411}
\end{figure}

In Fig.~(\ref{ab120}), we consider two possibilities: (i) the
standard case with $w^0_e:w^0_\mu:w^0_\tau=0.5:1:0$ leading to
$\bar{R}=2.53$ (see Fig.~\ref{ab120}-a); (ii) the case of stopped
muons with $w^0_e:w^0_\mu:w^0_\tau=0:1:0$ yielding $\bar{R}=3.20$
(see Fig.~\ref{ab120}-b). From these figures we observe that with
a precision of $\Delta R/\bar{R}=7\%$, these two scenarios can be
easily discriminated. These two can also be discriminated from the
scenario in which the neutrino production mechanism is $n\to p e
\bar{\nu}_e$ ({\it i.e.,} $w_e^0:w_\mu^0:w_\tau^0=1:0:0$). When we
restrict the analysis to $w^0_\tau=0$ ({\it i.e.,} the case
without exotic neutrino properties) from these figures we observe
that the measurement of $R$ stringently constrains
$w_e^0:w^0_\tau$ which in turn sheds light  on the production
mechanism. However, once the assumption of $w^0_\tau$ is relaxed,
a wide range of $w^0_e:w^0_\mu:w^0_\tau$ can be a solution. For
example, the exotic case of $w^0_e:w^0_\mu:w^0_\tau=0:0:1$ leads
to the same value of $\bar{R}$ as the stopped muon scenario.

The input for Fig.~\ref{ab411} is the same as that for
Fig.~\ref{ab120} except that in Fig.~\ref{ab411}
$w^0_e:w^0_\mu:w^0_\tau=4:1:1 \,.$ For this flavor ratio, the
central value of $R$ is $\bar{R}=1.96 \, .$ Notice that with
$\Delta R/\bar{R}$, this exotic flavor ratio can be discriminated
from the two standard cases that we have mentioned. That is
$w^0_e:w^0_\mu:w^0_\tau=0.5:1:0$ or $w^0_e:w^0_\mu:w^0_\tau=0:1:0$
are not solutions for $R=1.96(1\pm 7\%)$\,.

%

\section{Conclusions and Discussions\label{conclusions}}

Under the assumption that the initial flavor ratios at the source
were $w_e^0:w_\mu^0:w_\tau^0=1:2:0$, we have studied the
possibility of deriving $\theta_{13}$ and/or $\delta$ from cosmic
neutrinos taking into account various uncertainties.

ICECUBE and other neutrino telescopes that are going to collect
neutrino events in the energy range $100~{\rm GeV}<E_\nu<100~{\rm
TeV}$ will be sensitive only to two types of events; {\it i.e.,}
shower-like and $\mu$-track events. The ratio of these two, $R$,
yields only one piece of information on the mixing parameters.
Under the assumption of CP conservation, $\cos \delta=\pm 1$, we
have discussed the possibility of extracting $s_{13}$ from the
measurement of $R$. We have found that for $\cos\delta =1$, the
sensitivity of $R$ to $s_{13}^2$ is very mild. For $\cos
\delta=1$, the derivation of $s_{13}$ from $R$ would require
measurement of $R$ with a precision better than 2~\% which does
not seem achievable. In the case of $\cos \delta=-1$, as
$s_{13}^2$ varies between zero and the present upper bound, $R$
changes by $10$~\% which is in principle resolvable by ICECUBE
\cite{Beacom}.

We have found that a 10~\% uncertainty in the energy spectrum (to
be precise, 10~\% uncertainty in the spectral index in
Eq.~(\ref{power-law})) is a major source of error in the
derivation of $s_{13}$. However, for $s_{13}^2>0.02$, the
solutions with $\cos \delta=1$ and $\cos \delta=-1$ can be
distinguished despite a 10~\% uncertainty in the spectral index.
We have also studied the effect of a deviation from the power-law
spectrum. Our conclusion is that in order to derive $s_{13}$ from
$R$, a precision better than 5~\% in the measurement of the energy
spectrum is required.

We have also studied the effects of the variation of the neutrino
to anti-neutrino ratio on $R$. We have found that when
$\mathcal{N}_{\bar\nu_e}/\mathcal{N}_{\nu_e}$ (see
Eq.~(\ref{power-law}) for the definition) varies between 0 and 1,
$R$ changes by 5~\% which is comparable to the effect of $s_{13}$.
Unfortunately with the current techniques, it is not possible to
measure $\mathcal{N}_{\bar\nu_e}/\mathcal{N}_{\nu_e}$ in the
energy range $100~{\rm GeV}<E_\nu<100~{\rm TeV}$ so this source of
uncertainty cannot be eliminated by measurement and one should
rely on the models to predict the value of
$\mathcal{N}_{\bar\nu_e}/\mathcal{N}_{\nu_e}$.

The uncertainty in the neutrino nucleon cross section,
$\sigma_{\nu N}$, induces relatively large uncertainty in the
evaluation of the shower-like and $\mu$-track events. However,
when we take their ratio, the uncertainties cancel each other out
so the derivation of the neutrino parameters from $R$ does not
suffer from the uncertainty in $\sigma_{\nu N}$.

We have then studied the possibility of deriving $\delta$ from the
cosmic neutrino flavor composition assuming that $s_{13}$ will be
measured by other experiments with a reasonable precision. Since
the dependence of the oscillation probabilities of the cosmic
neutrinos on $\delta$ is through $\cos \delta$, there is a
symmetry under $\delta \to 2\pi-\delta$. On the other hand, the
sensitivity of the CP-odd combination $P(\nu_\mu \to
\nu_e)-P(\bar{\nu}_\mu \to \bar{\nu}_e)$ (which is proposed to be
measured by  neutrino factory or superbeam facilities) to $\delta$
is through $\sin \delta$ and would therefore suffer from a
degeneracy under $\delta \to \pi -\delta$. To resolve the latter
degeneracy, it is suggested to employ various baselines
\cite{baseline} and/or study the energy dependence of the
oscillation probability \cite{energy-dependence}. Derivation of
$\cos \delta$ from the cosmic neutrino flavor composition can be
considered as an alternative method to solve this degeneracy. We
have studied the effects induced by the error in the mixing angles
($\theta_{12}$, $\theta_{23}$ and $\theta_{13}$) and the
measurement of $R$ as well as by the uncertainties in
neutrino-antineutrino ratio and the energy spectra. We have found
that even with a precision of 1\% in $R$, CP cannot be
established. That is even for the maximal CP-violation
($\delta=\pi/2$), $\delta=0$ cannot be ruled out. However, in this
case, $ \delta=\pi$ can be excluded provided that $\Delta
R/\bar{R}$ is reduced to less than 1\%.

Within the SM of the particles, various mechanisms can deviate
$w_\mu^0/w_e^0$ from 2 but within the energy range $100~{\rm
GeV}<E_\nu<100~{\rm TeV}$, $w_\tau^0$ still remains much smaller
than $w^0_e$ and $w^0_\mu$. The conditions in the source of cosmic
neutrinos are so extreme that the beyond SM properties of
neutrinos can play a role to significantly distort the initial
flavor ratio of the cosmic neutrinos. In particular, a nonzero
$\mu_{\tau e}$ or $\mu_{\tau \mu}$ transition moment close to the
present bound can lead to $w^0_\tau \sim w^0_\mu \sim w^0_e$. In
the second part of the paper, we have relaxed the assumption
$w_e^0:w_\mu^0:w_\tau^0=1:2:0$ and have studied the possibility of
extracting the initial $w_e^0:w_\mu^0:w_\tau^0$ from the cosmic
neutrino data. We have found that with a precision of $\Delta
R/\bar{R}=7\%$, one can discriminate between the standard picture
with $w_e^0:w_\mu^0:w_\tau^0=1:2:0$ and the stopped muon scenario
with $w_e^0:w_\mu^0:w_\tau^0=0:1:0$. When based on theoretical
prejudice, we restrict the analysis to $w_\tau^0=0$, we find that
$w^0_\mu/w^0_e$ can be constrained with reasonable accuracy but
relaxing $w^0_\tau=0$ will open up the possibility of different
solutions.

We have also enumerated a number of other effects that can be
comparable to that of nonzero $s_{13}$ but have been overlooked in
the literature. Calculating these effects requires the knowledge
of details of the detector and the shape of the neutrino spectrum.
Uncertainty in this knowledge will lead further uncertainty in the
determination of $\delta$ and $\theta_{13}$. As we demonstrated in
the present paper, even in the absence of these effects, the
uncertainties are too large to allow for the determination of
$\delta$ and $\theta_{13}$. Potential uncertainties in these
effects will further confirm this conclusion.  For the purpose of
establishing substantial deviation of $w_e^0:w_\mu^0:w_\tau^0$
from $1:2:0$, these effects  have to be taken into account however
the uncertainties in the evaluation of these effects are not
expected to be so large to change our positive conclusion in the
second part of this paper. Throughout this paper, we have assumed
that the propagation of the cosmic neutrinos from the source to
the detector is governed by the standard oscillation formula. The
effects of a deviation from the standard oscillation probability
will be presented elsewhere.

\section*{Acknowledgement}

The authors are grateful to R.~Gandhi, M.~H.~Reno, M.~M
Sheikh-Jabbari and A.~Yu.~Smirnov
 for useful discussions. Y.F. would like to thank ICTP
where part of this work was done for support and kind hospitality.
A. E. would like to thank ``Bonyad-e Melli-e Nokhbegan'' for
partial financial support.

\appendix
\section{Calculation of $f(E_\tau,E_\mu)$\label{function}}

In this section we calculate the function $f(E_\tau,E_\mu)$ which
is the probability density of the emission of a muon with energy
$E_\mu$ in the decay of a $\tau$ lepton
($\tau\to\mu\nu_\mu\nu_\tau$) with energy $E_\tau$ (see
Eq.~(\ref{f-defintion}) for the definition). In the rest frame of
$\tau$, the partial decay rate of an unpolarized $\tau$ is given
by the following well-known formula (see \cite{griffith}) \be
\frac{1}{\Gamma'}\frac{d^2 \Gamma'}{dE_\mu' d\Omega'}
=\frac{12}{\pi
m^3_\tau}\left(1-\frac{4E'_\mu}{3m_\tau}\right)E'^2_\mu, \ee where
the effects of $m_\mu^2/m_\tau^2\ll1$ are neglected. Quantities in
the rest frame of the decaying $\tau$ lepton are denoted by a
prime. In the rest frame of the $\tau$ lepton,
$0<E'_\mu<m_\tau/2$. The number of emitted muons in certain
direction within the solid angle $d\Omega'$ and with energy in the
interval $[E'_\mu,E'_\mu+dE'_\mu]$ is a Lorentz invariant
quantity: \be \frac{1}{\Gamma}\frac{d^2\Gamma}{dE_\mu
d\Omega}dE_\mu d\Omega=\frac{1}{\Gamma'}\frac{d^2\Gamma'}{dE'_\mu
d\Omega'}dE'_\mu d\Omega'. \ee From this equality, we obtain
\be\label{function1} \frac{1}{\Gamma}\frac{d^2\Gamma}{dE_\mu
d\Omega}dE_\mu d\Omega=\frac{12}{\pi
m_\tau^3}\left[1-\frac{4}{3m_\tau}\gamma (1-\beta \cos
\theta)E_\mu \right]\gamma(1-\beta\cos\theta)E^2_\mu dE_\mu \sin
\theta d\theta d\phi,\ee  where $\gamma=E_\tau/m_\tau$ and
$\beta=\sqrt{1-1/\gamma^2}$. The $z$-axis is taken along the
direction of motion of $\tau$. The quantities $E_\mu$, $\theta$
and $\phi$ take values in the following intervals \be 0\leq
\phi<2\pi
 ,\qquad  0<E_\mu<\frac{E_\tau}{2}
(1+\beta) ,\qquad 0\leq\theta\leq \theta_{max}  \ee where  \be
\label{theta}
\theta_{max}=\arccos\left[\max\left\{\frac{1}{\beta}\left(1-\frac{m_\tau}{2\gamma
E_\mu}\right),-1\right\}\right]. \ee By integrating over $\theta$
and $\phi$ in Eq.~(\ref{function1}), we obtain \be \label{f1}
f(E_\tau,E_\mu)=\int_{\phi=0}^{2\pi}\int_{\theta=0}^{\theta_{max}}
\frac{1}{\Gamma}\frac{d^2\Gamma}{dE_\mu d\Omega} \sin\theta
d\theta d\phi .\ee  In the limit $\beta \to 1$ (or equivalently,
$\gamma\gg 1$), \be f(E_\tau,E_\mu)\simeq
\frac{5}{3E_\tau}-\frac{3E_\mu^2}{E_\tau^3}+\frac{4E_\mu^3}{3E_\tau^4}.
\ee

\end{document}